# The End of effective Law Enforcement in the Cloud? – To encypt, or not to encrypt

Steven Ryder

Europol
Steven.ryder@europol.europa.eu

Nhien-An Le-Khac
School of Computer Science
University College Dublin
Ireland
{an.lekhac}@ucd.ie

*Abstract*—With an exponentially increasing usage of cloud services, the need for forensic investigations of virtual space is equally in constantly increasing demand, which includes as a very first approach, the gaining of access to it as well as the data stored. This is an aspect that faces a number of challenges, stemming not only from the technical difficulties and peculiarities, but equally covers the interaction with an emerging line of businesses offering cloud storage and services. Beyond the forensic aspects, it also covers to an ever increasing amount the non-forensic considerations, such as the availability of logs and archives, legal and data protection considerations from a global perspective and the clashes in between, as well as the ever competing interests between law enforcement to seize evidence which is non-physical, and businesses who need to be able to continue to operate and provide their hosted services, even if law enforcement seek to collect evidence. The trend post-Snowden has been unequivocally towards default encryption, and driven by market leaders such as Apple, motivated to a large extent by the perceived demands for privacy of the consumer. The central question to be explored in this paper is to what extent this trend towards default encryption will have a negative impact on law enforcement investigations and possibilities, and will at the end attempt to provide a solution, which takes into account the needs of both law enforcement, but also of the service providers. It is hoped that the recommendations from this paper will be able to have an impact in the ability for law enforcement to continue with their investigations in an efficient manner, whilst also safeguarding the ability for business to thrive and continue to develop and offer new and innovative solutions, which do not put law enforcement at risk.

*Keywords—;*

## I. INTRODUCTION

Cybercrime has been as old as the invention of the internet, and arguably even predates that [1]. Where there is opportunity, there is crime. And where there is technology available which can assist crime, it will be used for such, even if it is not its (primary) purpose [2]. The ever increasing advances in computer sciences and technology have always been increasing the challenges faced by law enforcement, not only from a technological aspect, but also from the required imagination of how certain technology may be used, and the understanding that nothing is necessarily what it seems.

The next advancement and almost routine step is encryption, which in an increasing amount of laptops, computers or storage devices is already enabled by default [13] [14], to the frustration of law enforcement [15]. Encryption specifically can apply selectively to any and all storage devices either in their entirety or selected areas or even only selected files or folders. However, while encryption is indeed noticeable, further free tools are available, with minimum user capabilities, to create hidden and encrypted areas of storage, which on initial inspection, and unless actively being looked for, will be difficult to even detect (the concept of plausible deniability) [16]. In essence it also allows the entry of a second password, similar to a panic code on an alarm – which gives the impression of having deactivated, while in reality it received a different command.

Recently, with the development of Cloud computing platforms, cloud-based applications and new capabilities are emerging daily and bringing them lower cost of entry, pay-for-use processor and data-storage models, greater scalability, improved performance, ease of redundancy and improved of business continuity. Hence, more and more users, organisations select cloud computing as a solution of their IT platforms and services. However, this also raises challenges for digital forensic investigators. This could mean that for example a given collection of evidence such as illicit material is either stored on the servers of a cloud provider, or maybe even across multiple cloud providers. Bearing in mind as well though that the country of operation of the cloud provider does not necessarily allow conclusions about the geographic location of one or more of its servers, which could be hosted in multiple countries – let alone reflecting on the possibility that the original cloud provider himself has subcontracted its services to other providers [18] [19]. This means in practical terms that the illicit material stored by the suspect is literally scattered in the clouds, and for all intents and purposes for gaining access to it, it may as well be from a law enforcement perspective, who would possibly be required to identify the various locations, issue specific and separate requests for mutual legal assistance, and then hope for a swift response from the various jurisdictions contacted [20]. The challenges also include in order of the discovery of a suspect / suspected activity, first of all the identification of the use of cloud services for storage. This could be an obvious one such as the use of a Dropbox account which is prominent on the desktop, or a Google Drive icon, or even those provided immediately by the manufacturer of the device6 – but could also be less obvious solutions which are not immediately considered – such as using an email account itself for storing data in for example unsent messages / drafts. Should the use of cloud storage be suspected, and a specific provider is identified, the next challenge is the location of any stored material and its identification. Should the suspect not cooperate, the challenges increase – and increase to such an extent that the non-cooperation by the suspect as regards the provision of passwords itself has been made a criminal offence in a very pragmatic manner in some locations, such as in England and Wales [21].

The essential aspect to be examined and discussed in this paper is the scope to which extent the application of standard law enforcement investigative techniques and procedures of gathering evidence by use of e.g. court orders is indeed a sustainable approach. This can be questioned by comparing the uniqueness of the service providers in question, and the often complex structure of the providers own network or platform, with those of other industry areas. A cloud provider may as mentioned have stored the material in one location in its control, or may have spread it across multiple servers, or have no control at all about its whereabouts based on further subcontracting of the storage. In these situations it is hard to see what the best course of action is [32]. We also propose a solution to tackle these issues. The rest of this paper is organised as follows: Section 2 shows background research in this area. We present and discuss on law enforcement requirements in Cloud forensics in Section 3. We describe and evaluate our approach in Section 4. Finally, we conclude and discuss on future work in Section 5.

## II. BACKGROUND

In this section, we discuss on the essential Requirements, Governance and Challenges of Cloud Storage Service Providers.

Cloud storage is a massive business area and the biggest difference is the scale to which they outsource their own services and act more of an intermediary, rather than a full-service provider using their own infrastructure. The second difference is equally the size of their average client, and whether their primary business is the provision of entire servers to individual clients using high amounts of processing and / or storage, or whether the primary income is derived from services to the public (i.e. individuals), with a high ratio of clients to server (in which case naturally the removal of a server has a much bigger impact in the number of clients affected) [43].

This brings with it a number of specific challenges, when taking into account the need or even ability for private companies to be able to preserve or even access their own logs or infrastructure for forensic purposes12. As the challenge is large enough already at present, to simply gain access if you own the entire infrastructure, let alone if you do not, as is the case when data storage is possibly further sub-contracted. Additionally, due to a large part presumably to the heightened public awareness concerning data retention and collection, following the revelations by Snowden, companies are actually profiling themselves by being unable to provide logs or forensic evidence to law enforcement - see for example in this regard Apple, a market leader and a brand with a strong image as trend setters, publicly claiming that they are unable to comply with law enforcement request, as they cannot access users data (anymore) [44] [45]. From a conceptual level, this is achieved by basing the encryption key on the user's individual pass code, making every encryption unique. This naturally brought strong criticism from the FBI, and various sources of law enforcement, but possibly contributed to Apple's increased popularity (e.g. a rise of 67 percent from 2014) [47]. As a result, if this is the new standard that is being set, then other companies will need to follow suit – as Google has also recently announced to follow up this approach with one if its next Android version [48]. One could say that therefore this trend has already spilt over as a specific sales point towards other providers of cloud services. This indeed creates the question of what, other than binding legislation, would bring companies to cooperate, if the public attitude was pro-privacy, and against law enforcement, and further at the same time companies encrypt their own records beyond even their own retrieval possibilities, or don't keep them at all. Especially as in the given circumstances even then the risk or threat by law enforcement of seizing servers or discontinuing their business is fruitless, as law enforcement would probably equally be unable to gain access to the data in a meaningful manner, hence making the entire seizure pointless in advance. Even stronger, where such a seizure would take place in any event, the businesses affected may well be in a strong position to claim damages due to the foreseeability of the damage and at the same time the non-existent likelihood of the success of any data recovery from the seized equipment.

### A. Requirement for encryption

The utilisation of cloud storage commenced with business usage. As more and more businesses outsourced their data storage as well as in many cases business processes and running of software, the more the question of the security of the data was raised. This was initially a rare occurrence, but has increasingly become almost a daily occurrence. Additionally, while data originally being processed was limited to internal data, with the rise of commercial transactions taking place over the internet, and the ability to process credit cards off-site, the concrete value of the data became apparent, as did the consequences of a data breach, moving from an embarrassment (if anybody even found out) to actual significant financial damage. As a result, the processing of card details, is it under the data protection framework requiring appropriate measures to ensure data security, or be it under terms and conditions applicable to merchants from their banks, require in essence encryption.

All recent data breaches which have reached the public nature, be it the JP Morgan breach [87], the PS Network [88], Sony Pictures [89], adult dating websites such as Adult FriendFinder [90] and the Ashley Madison website [91], as well as most recently the breach of The Hacking Team (which specialises in developing custom made malware for use by governments across the globe) [92], were severely criticised for not encrypting all of their data, such as passwords, user names, or financial credit card details which resulted in either the publishing of the documents or data obtained, or may well be currently available for purchase in certain fora, or even used in blackmailing attempts at the company or the users exposed [93]. As such, one could reflect if the courts may also seem willing to apply an accusation of negligence, for not having encrypted their data.

This is an ever increasing area of requirements, especially as hospitals move to store data in the cloud, allowing for easier information sharing, which contain the most sensitive data on a person, but equally explorations are ongoing for law enforcement and judicial authorities such as courts, to utilise cloud storage and cloud sharing. It goes without saying that a data breach of such an authorities cloud would cause outrage at plain text storage of such data.

Naturally, and an obvious question now is, to what extent can you differentiate between good and bad encryption? As by its very definition, and the objection to encryption from a law enforcement perspective, is specifically to allow the evaluation of whether or not something illegal may be ongoing. As a result, it seems paradox to call for a prohibition of encryption if at the same time, it is an essential requirement.

Equally, solutions such as what the author believes can be called pseudo-encryption, i.e. encryption, but with certain authorities possessing a master-key or backdoor, seem unsuitable as they would equally jeopardize the safety of all other material which is encrypted. It is also an unprecedented intrusion into the rights of all citizen, rather than a targeted approach intruding only on the liberties of one citizen, the suspect. Apart from the proportionality aspects, it would also face significant practical challenges [94], and most likely completely undermine any trust of citizens in government, and in the end lead to the development of new encryption strategies [95].

Overall, it can be held that encryption in a number of industries, including law enforcement, is not only best practice, but a business essential, without which no real secure environment required could ever be established in any event.

A final aspect to raise is that even though encryption on its own may be seen as dangerous for specific groups, such as law enforcement, it is one of the only efficient defences against an ever increasing number and styles of cyber-attacks. With the prevalence of software, exploits, (semi)public availability of tutorials, and proportionally low cost rental availabilities of foreign servers or even individuals themselves, it is impossible to ensure that a system which is designed to be accessible from any location across the globe via the internet, will at one stage not be compromised and data stolen. Encryption therefore is understood to be the last line of defence, and as such, it cannot be outlawed safely. And this is the final point about mandatory encryption being required – the biggest security risk to any system is actually the user. Ranging from poor password selections, to the level of general awareness and understanding of computers and the internet – the user is in essence the most resilient to positive change, while at the same time being the most important.

The number of examples available, from the average user unable to explain the difference between Google and the internet, or let alone between Google and Internet Explorer for example, to those that should know better abound mercilessly. Examples of the former, albeit indisputably non-academic examination and of a very colloquial nature are updated daily on Reddit's Tales from Tech Support [96]. Examples of those who should be aware of their responsibilities and implement best practice concern the leaving of default passwords on road signs, ATMs, and Point of Sale Systems (i.e credit card and PIN terminals) [97], as well as the latest hack of the Hacker group, who are themselves at the forefront of compromising systems in the name of various governments, who neglected to enforce proper security on their own systems. And the final example in this respect, if it is even needed to provide evidence in support of a common truth, are the number of YouTube videos demonstrating external parties taking over control of baby monitors which are monitored from the internet – and should have had parents provide their own password [98].

The merit of all these seemingly unrelated arguments is that in the end internet security depends on the wrong or inaction of a user. And the fact that for example industry such as apple and Facebook and to an extent also Google, are moving towards default encryption just emphasizes this point. They believe (rightly) that the consumer cannot be trusted to make the right choices, so they will be made for him. The same applies in principle to Windows, which if every user of Windows updated all of their operating systems as prompted and even done automatically for the consumer if the user just allows it, would be the biggest blow, and if not the extinction, of the number of botnets in operation today (notwithstanding the number of illegal or compromised Windows OS originally installed, and which for example in China, where an estimated 90% of Windows users do not own a legitimate copy, and are reluctant to upgrade, as they are unsure whether they can bypass the security of the then upgraded system [99]).

At present, the legal framework governing encryption is not in existence as regards the prohibition thereof. The only legal framework in existence that governs encryption, actually demands encryption, rather than prohibiting it, as stated above.

Further, encryption can be seen as the last line of defence, and the setting of it as a default is, whilst a clear manner in which a decision is made for the user, without his real involvement, a necessary aspect to protect the user from his own actions. The outlawing of encryption, while providing a boost to law enforcement's ability to investigate, would be more than outweighed by the additional number of investigations they would have to launch to deal with the floodgates on cybercrime having opened, combined with the increase of financial and economic loss – and quite possibly the end of e-commerce itself. This in general seems to have been accepted by law enforcement as well, as will be discussed next as regards the possibilities open to law enforcement.

*B. Encryption Techniques and Brute Forcing*

A brief and generic overview of encryption and password security is helpful to place the below concerning the needs for law enforcement and the proposed solution into context. There are numerous papers, from articles to PhDs dedicated to how encryption works on a technical level. Upon to [102], the encryption has no real relevance on the ability to intercept or gain access to the material. What it does mean however, is that a 'key' is required to decrypt the content. As the key is user defined, it is now necessary to gain access in the absence of the user's cooperation. The main approach can be called 'brute force'. This is essence requires trying various possibilities one after the other. Generally, this follows the route of trying either a database of popular passwords (see e.g. 10 million passwords publically available collated from previous password breaches [103]) or simple dictionary words first, before moving to moving through the various character possibilities.

Additionally, and upon initial reflection, with dedicated computer composition allowing for 3.3 billion guesses per second [106], it appears only a question of time until any and every password is found. This is indeed literally correct, yet the exponential factor of the time increase is often neglected, based on the presumed length of the password. For example, a 5 character password, containing an uppercase letter, a lowercase letter, a number and a symbol, will take approximately 3 months when a brute force attack is undertaken (ignoring the use of dictionaries or previous passwords), at 1000 guesses per second, which is the expected volume in an online attack [107]. An offline attack on the same password could reach one hundred billion guesses per second, and a dedicated attack on sophisticated high end equipment could even reach one hundred trillion guesses per second – in both cases, arriving at the 5 character password in less than a second.

However, building on the exponentiality, and on the presumption that the composition of the password contain the four different types (i.e. at least one upper- and lowercase letter, a number and a symbol) we require a password length of ten, in order to challenge the dedicated attack scenario at 100 trillion guesses per second – which would then take about a week. For general context, at this stage already, the online attack (at 1000 guesses per second) would take approximately 19 million centuries to test all possible combinations, and even at 100 billion guesses per second, it would still take up to 19 years. If one more digit is added to the password and the length reaches 11, the dedicated attack would still take up 2 years, but if one more digit is added, with a total length of 12, the time it would take to test all possible combinations would be around 200 years, even at 100 trillion guesses per second. For comparison, the table below provides for an overview of the various scenarios:

With the awareness of these figures, and despite the fact that even though we have been spending years "teaching humans to learn password which are easy to crack but hard to remember" [104], if humans to further that quote do change their passwords to non-dictionary, and non-routinely used passwords, and they move to 12 characters, the ability to brute force the passwords is not a reality.

III. PROBLEM STATEMENT-LAW ENFORCEMENT REQUIREMENTS

As always, law enforcement, arguably due to the nature of being a public authority, and usually always outmatched as regards financial resources and priorities, is always a step behind in the development of new technologies and their usage for committing or concealing crime, albeit also this is changing. See here for example also the interaction of LEA with private companies in the development of new products, e.g. SKYPE and Microsoft, but also compared with pre- and post Snowden – with Apple as a global market leader providing default encryption which it itself states it cannot unlock. As business needs are driven by perceived consumer demand, this is a remarkable step which will likely be followed by other market leaders. That genuine law enforcement needs are hindered to an extent that fears of privacy intrusion outweigh the need can also be seen as a public backlash.

Overall, the law enforcement authorities, in order to effectively combat cybercrime specifically, require the ability to translate traditional methods of investigations, such as surveillance, eavesdropping, wiretaps, or intercepts, to be applicable to modern means of communication, i.e. email, instant messaging, chat rooms, and even cloud services themselves, for example the use of web based email systems for storage. While most countries are in a position to apply existing national legislation in analogy, and some have specific legislation to assist this (especially those party to the Council of Europe Convention, as elaborated above), what is lacking is the ability to implement these measures from a technical perspective. Not only will the possibility exist for most users to encrypt all communication with free open source tools, but more importantly, and somewhat uniquely, especially as regards cloud forensics, is the fact that the law enforcement authorities depend to a very large degree on the provider having useable logs and tracking integrated into their own system platforms from an operational systems perspective / cloud management aspect. As without this third party data, there is no record of any crime being committed. And this dependency lies at the very heart of the potential conflict between law enforcement and cloud service providers.

The main purpose of computer forensics is to gain, as with any other piece of evidence, as much information as possible from an object. The unique aspect of computer forensics relates to mainly two types of scenarios that require a degree of skill set or specialized software. One is the occasion whereby a computer system or any storage device needs to be assessed for deleted files or hidden or encrypted content, and ideally recovered, the other can probably be seen as occasions whereby a live system or network needs to be investigated and activity on a server for example reconstructed, and matched with potentially identified IP addresses or the like. Both scenarios would then also, depending on the investigative requirements, allow for social network analyses, of who is sending who emails, which contacts are in instant messaging tools, what chat history or logs are available, etc [130].

For cloud storage purposes, there are in essence two possibilities which open up, even though they may be generalized. Police may be dealing with a cooperative suspect, who provides them all the details about his account, from the provider, his username and password, and the location of any illegal material. In such a case, the forensic task is not as great, as armed with this information, the question is only concerning the recovery and the documentation of that access.

It becomes significantly more challenging, when dealing with a suspect who is believed to have access to cloud services, but whose devices are fully encrypted, and neither a specific company, nor a user name is known – let alone a password. In these scenarios, it is a painstaking process, which may even end fruitless – dependent on the level of encryption or traces of his internet activity from the Internet Service Provider itself, if it even keeps those records [131].

In such scenarios, if at least a company is identified that provides the service to the suspect, and then the next step would be to seek its cooperation.

In conclusion, the needs for law enforcement which should be contained in any feasible solution take into account the operative need to be able to reconstruct events from a historical perspective, to be able to identify and gain access to certain accounts, as well as be able to obtain a copy of the data stored, combined with ideally a log of all activity pertaining to that account.

However, from a more internal perspective, the largest requirement would be a dramatic increase in training and awareness – as here it can also be held that while police can and should be seen as an average reflection of members of society, this would also imply that the general awareness of cyber security and cybercrime, and the required user behavior in light of this, may also be somewhat lacking.

Finally, an ability to obtain all the data required for their investigation, without the need to depend on external or third parties would be ideal. This would however imply receiving a cooperative suspect in all cybercrime cases, and this seems a slipper route to take. However, with a number of analogies, one of the solutions should consider this possibility of well of enforcing cooperation, also by the suspect himself.

IV. PROPOSED SOLUTION

Solving this problem, which is not exclusively related to the small snap-shot of cloud computing, but has significantly further reaching consequences as concerns general questions of moral values of a society, is not a simple undertaking. Rather, it requires a refined approach, and it also needs to reflect the fact that cloud computing specifically is cross-border, or most likely even without locatable physical locations – meaning a degree of global jurisdictional applicability is required [142].

In the following few chapters, it is intended to assess proposed solutions for their feasibility and likely impact on law enforcement, as well as service providers, and to develop from the various proposals as well as the various downsides, a new approach.

It can be held that there are complex matters to be taken into account as regards all online crime, but specifically also the usage of cloud storage. Mainly in literature this aspect has been approached from the point of view of law enforcement and judicial complications, but not really from a business need perspective. While I suppose a reason for this lays in the rather limited amount of cloud storage providers over the past, and that being mainly a business oriented market with clear dominating powers, which is changing fast. As described previously, the market is becoming less and less dominated by main players, and creates a lot of room for localized cloud storage markets, including for personal use.

This means that in general terms, the larger the company, and the more equally law enforcement is dependent on their cooperation and good will, the clearer the informal arrangements between the two – and law enforcement enquiries or investigations no longer pose a business continuity threat to the providers. This changes, if the market is flooded with new companies, which either do not wish to take a cooperative approach with law enforcement (while however still being strict about their content and services, but the lack of cooperation being based on privacy concerns and a strict adherence to legal procedure and process), or which are simply unaware of the impact law enforcement action could have on their business.

This by definition makes the entire question of this thesis more relevant again, as how should the market behave? This has never really been truly examined. This is also why in the below proposals, I will address and provide proposals focused on law enforcement and judicial aspects, but also provide solutions based on the business side of commercial cloud storage providers.

*A. A clear legal environment in which the business operates*

Taking into account the previous aspects raised as concerns the questions of jurisdictional applicability, any business needs to know the legal environment in which it operates. However, there are specific challenges for a cloud storage provider, or also a cloud service provider in more general terms, based on the geographic location of its actual storage facility. It would appear from the various terms and conditions applicable to the provision of services, that the majority of the providers nominate a court / jurisdiction applicable to the contract of service. However, this contract cannot override national law. And as such, a storage provider with multiple places of business, which most storage providers are, by virtue of the irrelevance of the physical location of its customer, will have to take into account multiple jurisdictions. As much as possible, this should therefore be avoided, and a degree of protection needs to be afforded to service providers, to know the terms of their engagement – and if need be, for example, have the liberty to exclude customers based in jurisdictions where the terms of engagement are not acceptable to the provider.

A feasible solution therefore must allow a business to make an informed decision on its operating environment, even if it does operate in a field freed to a large extent from geographically meaningful boundaries.

*B. A clear legal environment for its interaction with law enforcement*

A more specific subsection of legal clarity comes as regards the need to interact with law enforcement, and more specifically, what the possibilities for law enforcement are in that jurisdiction. It can not be the case that a business, when assessing one of its biggest and mission critical risks, has to rely on policy decisions in force at the time, based on the discretion of a judge able to issue a warrant, or the practice of the respective prosecutor. If it were subject only to the question of individual preferences and discretion, rather than concrete and specific legislation, it would, if wise, either opt not to operate in that jurisdiction, or to see itself forced to cooperate in an anticipatory manner, for fear of losing the business.

As such it needs to be clear, and that can only be done by explicit legislation, applicable to the specific sector, what can and cannot be done. Relying on the interpretation of legislation in force significantly before the emergence of the various technologies, the internet itself, or as a result of ill-informed legislators (while having good intentions), creating laws which have no real direct bearing on the process.

*C. A clear ability to maintain business continuity*

Apart from the technical causes that may cause system outages, be it from electronic causes, events of nature, or a pipe bursts, etc., there should be no other threat emanating towards its business continuity, and especially not from the side of law enforcement. This all pre-supposes naturally that the services offered by the business are indeed legal, and that there is no criminal activity conducted by it, or condoned / supported / encouraged by it. However, as regards the concerns of business continuity, it cannot be the case that the suspected engagement of one of their customers is an inherent business risk to the provider.

*D. A clear ability to provide services to clients, without being responsible for the content*

Any responsibility for the content stored on the cloud provider's systems cannot be made the business's responsibility. The situation is clearer as regards for example internet service providers, telephone companies – infrastructure providers in general. The moment the cloud provider is by default responsible for matters hosted on its servers, will be the moment that the cloud will no longer be able to be used by businesses, law enforcement, or any other person with a legitimate need to protect their data from third parties.

This is therefore an essential requirement, as equally from a provider perspective, the amount of staff it would take, as well as possibly skill, to review every bit of data stored in their cloud, would make the entire business model collapse.

*E. A clear relationship of trust with the customer*

The cloud provider needs to be able to assure the client that his data will be safe, inaccessible to third parties or the provider itself, and that the client can entrust his data to the provider, and see it in essence as an extension of his own desk or living room or garage – wherever else he may have stored the data otherwise.

If this trust is violated or not present to begin with, the company will not survive. And especially the possible pressure of advance or proactive compliance with law enforcement requests, for example those not supported by a warrant, damage such trust. Equally, data breaches and security breaches damage it – but not to a massive extent, if the data is encrypted.

*F. Summary of needs of Storage Providers and Law Enforcement*

It is easily summarized that the needs of a cloud provider (from a non-technical aspect, but limited to the scope of this paper), require clarity – simply and foremost: clarity on the jurisdictional aspects, clarity on the abilities of law enforcement, and clarity on questions of liability. So any feasible solution must contain this one core ingredient. The outcome of the clarity may not be necessarily in the best interest from a business perspective, but at least then an informed decision can be made about the establishment of a branch or headquarters, or the provision of services to a specific country or area.

From a law enforcement perspective the main requirements are almost identical, in the sense of requiring also a need for a clear legal framework in which to operate, and an ability to gain access to data it requires in a non-time or resource consuming and efficient manner, while naturally having a lawful order from the court to do so.

*G. Way forward*

We, upon reflection and the acknowledgement of all of the above discussed aspects, comes to the following proposed solution. This is primarily based on the realization of two aspects, namely that laws need to be seen to function, in the absence of which no meaningful policing can take place, and secondly, that with the advent of the internet, and the Snowden revelations just being a catalyst for this belated realization, a significant change has entered into the expectations of society, namely that a small part of our lives is governed by near anarchy, where anything goes and no one is accountable to anything, for anything or to anyone. This refuge from the daily laws and regulated life worked arguably fine, but has now become such a dominant part of our lives, that the status quo is not maintainable, as much as it may be desirable by some parties.

We believe firmly in the rule of law and the role of the law, albeit from a legal realist point of view if one takes into account the legal philosophy behind the entire construction.

In full awareness that the common law system is as far as the main focus concerns the European Union, is rather in minority, the principle nonetheless must remain valid for any legal system that seeks to be respected and complied with.

The question of forensic evaluation or obtaining of access to cloud services for law enforcement is a necessity and question which was not really an issue or even perceived challenge a decade ago. To date however, it has become a reality, and law enforcement has proceeded to attempt to apply their existing tools and abilities, and come to the realization, that in all aspects, it seems counterproductive (by alienating the service providers on whose cooperation they depend), expensive (forensic in and external evaluations and data recovery, storage of seized devices), and non-timely (taking into account backlogs). What is required is therefore an acceptance that the virtualization of near everything, at present, as well as thinking slightly ahead of the buzz words of the Internet of Things, is a challenge for law enforcement that does not possess the skill set or the ability to obtain what it needs – and not due to lack of skill or financial involvement.

Rather this is due to the fundamental nature of things. In the absence of a totalitarian supervision and monitoring of all activity of all citizens, police will be unable to proceed to gain access to what is needed. And here is an essential aspect when talking about what the police need to gain to conduct their investigation: If the police have a need, which involves obtaining something that requires a violation of an individual's right, i.e. the application of coercive powers, the police attend the court. The court then issues an order – which is binding on anyone it is directed at – such is the basic premise of our legal systems. This applies in civil disputes as much as in criminal cases, the court is the highest and only authority in the land which all citizens must obey.

The unique aspect here, and this is the key realization, is that to date, there is nothing comparable that the court can authorize the police to obtain, which the police cannot obtain themselves by force or other means in the obtaining of their goal. As a suitable analogy, if police apply for a warrant or order to search a house, in the lack of compliance of the owner or tenant, they can force their way in, within a reasonably short time frame. The same applies to any vault or the like found on the premises or subject to a warrant. What it does not apply to is the gaining of access to encrypted information. While methodologies exist to break encryption, dependent on the type of encryption utilized and the user chosen complexity of passwords (and here police may gain a break, considering the type of standard passwords chosen, but that is beside the point), they may take a disproportionate amount of time (up to multiple years if a secure password was chosen, and the only approach was brute force). The same, especially as regards to cloud storage and the connection to the storage, is applicable to live systems. Once a live system is shut down, the connections active at the time, as well as the probably unencrypted state of the hard drive(s), are likely to have been lost, possibly forever.

This means in essence that compared to any other coercive measure, ranging from highly intimate intrusions into a persons physical integrity like forced cavity searches, provision of medication to force a person to be sick, if they are suspected of having evidence in their stomach, such as drugs, to the more mundane of being ordered to provide a DNA sample, this is one of the only court orders available against a suspect, which the police cannot enforce. So what is they way forward when focusing on a valid order of a court, which the police cannot force compliance with or obtain by force on their own? In all other instances, for example in civil cases, if a court order is not followed or acted upon, or even when individuals are compelled to give testimony but refuse, what are the options of a court? To hold the individual in contempt of court. And this is a notion that has routinely been seen as an option in civil cases, with individuals being subject to considerable amount of time in court (at least in the US, the record stands at 14 years, based on the fact that he could comply with a civil order to transfer assets to his divorced wife, but simply refused to do so [144]). Most notably, in civil contempt cases, there is no principle of proportionality, which is the case in criminal contempt cases however, and hence the distinction is vital [145].

In criminal contempt matters, the subject matter of the current proposal made here, a principle of proportionality is essential, as are the existence of procedural rights for the subject, to for example submit a defense of impossibility. This follows very much the line taken by the United Kingdom's Key disclosure law.

As a general principle, this paper then proposes, taking into account:

- the uniqueness of the technological situation and problems faced,

- the need for service providers to be able to operate without undue interference from law enforcement investigations, both from a business continuity aspect as well as from the perspective of constantly being subject to court orders obliging them to provide password credentials or to decipher content stored on their systems

- the need for creating a situation whereby the law enforcement effectiveness is not dependent on third party compliance, and

- the accepted root concept of legality, the ultimate binding nature of court decisions and orders on every citizen, the core base upon which ultimately society is built

- the established context of contempt of court and its widespread use in civil cases

to create specific legislation on a national level, outlining the guidelines for the appropriate sentencing measures, to allow for individuals to be held in contempt of court for non-compliance with court orders to produce access to established storage deemed in his possession, while also ensuring that sufficient procedural safeguards are available. The reason this paper does not agree with the criminalization of the specific withholding of information, is that this is a determinate sentence. In cases of contempt of court, the accused continues to hold the key to his release, quite literally, and can be released at any time upon the production of the required access information. This retains also the spirit of the general principle of contempt, which is in essence seen as a coercive measure aimed at obtaining compliance. (And in German Law, the concept is known as "Beugehaft", which if literally translated, is the taking into custody ("haft") for the purposes of literally bending him over ("Beuge") , in the sense of submitting.)

Consequences of Encryption

Also in this respect, the problem is in essence removed, as the requirement to remove the encryption is burdened upon the suspect – not for ease and convenience, but because otherwise it would fundamentally inhibit any, let alone efficient, law enforcement.

As a consequence as well, the entire move of the industry towards encryption by default, which is still a large hindrance and obstacle for some law enforcement investigations, and significantly so it is presumed for intelligence agencies, is however largely inconsequential. Additionally, it would have the benefit of avoiding the reflexive need for cloud providers to move towards encryption, for fear of significant impacts on their business continuity.

Risks

The identifiable risk that appears most prominent is public reaction to such a proposal, as it would appear as a new concept, near Orwellian, to oblige a suspect to assist in his own conviction. If this were the case, any outrage would be justified as it would be incompatible with a functioning democracy and the rule of law. However, it is not that – rather, it is simply an application of existing rights of the court, in existence for as long as the existence of the courts themselves, and applied in order to achieve the implementation of a court order – which would otherwise be impossible to fulfill. Equally, what would be the point of the court ordering anything, what authority would a court possess, if its orders were in essence meaningless for lack of inability to enforce compliance?

Mitigation

Possible additional measures to complement the contempt specifications could be that criminal courts could refer the matter to a specialized court or tribunal, where specific membership and selection criteria as regards the judicial composition of it apply, ensuring only specialized judges are in a position to make informed decisions, and could maybe even include a member of an independent body dealing with civil rights.

These sorts of specialized courts already exist in a number of different areas, from employment tribunals, to special courts dealing with national security implications of cases, so an additional specialized court may well have merit, dependent on how strong the risk is perceived that ordinary judges holding criminal cases may be unable to deal and understand the complexities of cloud storage or the like.

What is also an additional benefit is the fact that in exchange for example for the shift in the application of these powers and the coherent national policies applicable to them when created, would mean that the necessity for example of obliging cloud providers to store all information about all of their customers, in decipherable format, is no longer as essential. While the establishment of a historic behavior may still be highly relevant, normal forensic evaluation should allow the establishment of such a picture when required. The point is therefore precisely this – it is not about having the suspect convict himself, it is rather the suspect giving the key to his house, in the absence of the ability of the Police to force entry, in the exercise of a judicial warrant.

## V. Conclusion and Future Work

The usage of cloud storage for criminal purposes is well known and will most likely not end. The eradication of crime is equally a noble cause, yet not one which seems a realistic prospect, and without any intention of entering into a philosophical discourse on human nature and societies. The fact that criminals and crime and their methods will continue to evolve in pace with and take advantage of technological development and innovations, is equally acknowledged. And the central aspect is that this is not necessarily an existential threat. It can equally be anticipated, that encryption, despite the strong concerns expressed within the US (where the Director of the FBI stated that he could not understand why companies "market something expressly to allow people to place themselves beyond the law") [147], most vocally within the EU by David Cameron [148], will be an increasing feature and become the norm.

As such, the proposed solution of moving away from expressing concerns about encryption, and moving towards individual responsibility and accepting the advancement of technology sometimes requiring an adaptation of law, should find less objection, as it specifically acknowledges the concerns about mass-surveillance and privacy of the individual, and makes the decryption an individualized order, rather than a mainstream standardization.

In this context of mass surveillance, a significant change of mindset and public perception of what normality is, has taken place, and will be hard to reverse. Trust, in its very essence, regardless of it being trust in public authority or government, is easily lost – and it takes a long time to restore, if it is possible at all. The manner in which generally the public has used the internet has been with a complete disregard to individual privacy protection, disregard for any terms and conditions and blind acceptance in general, including giving up their eldest child for free wifi-access [149], and those that were focused on security and privacy, including in their daily internet use, considered fanatics or conspiracy theorists. Even criminal activity is ongoing in a clear and transparent manner at large volumes, which is easy to detect and investigate, albeit time consuming, based on police being confronted with a "tsunami of filth" [150] (as concerns for example the United Kingdom, a recent report found that on average, 2 people a day are being convicted for possession of child abuse material, relating to approx 4.5 million images seized [151]). However, if the standard moves by default, even the high volume of for example child abuse material being collected and shared is moved into encryption, the work for law enforcement is dramatically multiplied.

Yet to stop the tide moving towards encryption, and specifically outside of the easy remit of law enforcement, will be a big undertaking – and for all intents and purposes, a non-productive approach in any event. As such, it is essential that law enforcement seek to cooperate as much as possible with private parties in the sector, and equally encourage them to embrace and accept their own responsibility in preventing the use of their systems for criminal activity (even if not necessarily a legal obligation, as can be seen from the zero knowledge approach adopted by a number of cloud providers, but rather their 'moral' obligation towards society).

It is only through creating this codependency, or rather co-responsibility, that action may be effective. As this however goes against the very point of encryption from a user – side, i.e. the concept of self-defense of the industry against being seen as law enforcement cooperative, it is going to require legislation to achieve in all likelihood. And specifically the type of legislation proposed in the solution in this paper would precisely provide the outcome of desire to all parties considered here – citizen, business and law enforcement.

Overall, the continuous development of cloud and virtualization services is permeating from a business exclusive aspect, to a mainstream consumer product. At the same time, encryption as demonstrated extensively above is equally advancing into a main stream feature. This has the potential to become a perfect storm for law enforcement investigations, at least as regards the currently deployed tactics. Equally, from a business provider's aspect, it is the ideal scenario, of being incapable of being blamed or held accountable for the abuse of its own services for criminal purposes, and that inability to detect it being in essence the fundamental business model.

There is little detectable self-interest in offering a service which offers less security on the basis that this will make work easier for law enforcement – especially if one is thereby in the minority in the industry, as less safety has never really been a selling point. And the fact that if a consumer uses one specific company for their cloud and virtualization services, and by reputation this makes them an honest, and transparent customer, is not really something to build a business case on. To date, the need for encryption, in its extreme form, bullet-proof hosting, or the use of Tor, has always had a publicly general connotation of the need to hide something, suggesting illegality. The service providers equally wished to remove connotations of illegality being a business model, and as such avoided that. With the switch however now, this is no longer the case – rather, there is no longer a contradiction between having nothing to hide and acting illegally. The black and white approach of having nothing to hide, and hence having no objection to being potentially watched, while still being prevalent in a lot of countries in the public, is not taken on board anymore by business – and hence there is no stigma attached to encryption and ensuring one's own privacy.

With the proposed (re-)introduction of the superiority and binding nature of the orders of a court to any citizen in its jurisdiction, society will ideally be able to reflect and realize that indeed, the notion of mass surveillance is at least in its current format, coming to an end – and that the intention pursued by law enforcement is not to know everything, about everyone at any time, but rather to be in a position to tackle specifically organized crime and terrorism, which should not be thwarted based on business decisions by private industry. Will the proposed solution make everyone happy? No. Will it hinder law enforcement in their investigations of volume based child abuse material investigations? Probably yes. Is this however a price to pay, to lead police to adopt and focus their resources on higher value targets, and to provide a feeling of privacy towards the citizen, with the ultimate result hopefully being a commencement of the rebuilding of trust towards law enforcement and government? Yes, the author feels strongly that this is the case, and is hopeful that the humble proposals made, the arguments advanced, and the overall value of this paper will see some discussion towards this goal.